\documentstyle{elsart}

\newcommand{\ice}[1]{\relax}
\newcommand{\ep}{\epsilon}

\newcommand{\bea}{\begin{eqnarray}}
\newcommand{\eea}{\end{eqnarray}}

\newcommand{\beq}{\begin{equation}}
\newcommand{\eeq}{\end{equation}}

\newcommand{\ba}{\begin{array}} 
\newcommand{\ea}{\end{array}}

\newcommand{\g}{\gamma}

\newcommand{\dsp}{\displaystyle}

\begin{document}

\begin{flushright}
\begin{tabular}{l}
  MPI/PhT/97-019\\
  hep-ph/9703278\\
  March   1997   
\end{tabular}
\end{flushright}

\mbox{}

\begin{frontmatter}
\title{
Quark Mass Anomalous Dimension to ${\cal O}(\alpha_s^4)$
      }
\author[Moscow,Munchen]{K.G.~Chetyrkin}
\address[Moscow]{Institute for Nuclear Research,
 Russian Academy of Sciences,
 60th October Anniversary Prospect 7a,
 Moscow 117312, Russia
                }
\address[Munchen]{%
Max-Planck-Institut f\"ur Physik, Werner-Heisenberg-Institut,
F\"ohringer Ring 6, 80805 Munich, Germany}
              
\begin{abstract}
We present  the  results of   analytic calculation of the 
quark mass anomalous dimension to ${\cal O}(\alpha_s^4)$. 

\end{abstract}

\end{frontmatter}


\noindent

\medskip

\medskip

\medskip

\noindent
\vfill
Corresponding author: K.G.~Chetyrkin, address: 
Max-Planck-Institut f\"ur Physik, Werner-Heisenberg-Institut, 
F\"ohringer Ring 6, 80805 Munich, Germany \\ 
e-mail:
chet@mppmu.mpg.de

\newpage

\section{Introduction}
The quark masses depend on the renormalization scheme and, within a
given one, on the  renormalization scale.  The dependence
on the latter is usually referred to as ``running'' and is governed by
the quark mass anomalous dimension, $\g_m$.

The  two-loop anomalous dimension is known for long \cite{gm2};
the three-loop result is also available from Refs.~\cite{Tar82,Larin:massQCD}.

The  quark mass anomalous dimension $\gamma_m$
is defined as 
\begin{equation}
\mu^2\frac{d}{d\mu^2} {m}|{{}_{g_{{\rm B}},
 m_{{\rm B}} }}
 = {m} \gamma_m(a_s) \equiv
-{m}\sum_{i\geq0}\gamma_{{i}}
a_s^{i+1}
{},
\label{anom-mass-def}
\end{equation}
where 
$a_s = \alpha_s/\pi= g^2/(4\pi^2)$, $g$ is
the  strong coupling constant. 
To calculate  $\gamma_m$ one needs to 
find the so-called quark mass renormalization constant, 
$Z_{m}$, which is defined as the ratio of the bare and renormalized
quark masses, viz.  
\beq
Z_m = \frac{m_B}{m} = 1  +
\sum_{i,j}^{0<j\leq i}\left(Z_{{m}}\right)_{ij}
\left(\frac{\alpha_s}{\pi}\right)^i
\frac{1}{\epsilon^{{j}}}
\label{Zm}
{}.
\eeq
Within the
${{ \mbox{{ MS}}}}$ scheme \cite{ms}
the coefficients $\left(Z_{{m}}\right)_{ij}$  
are  just numbers,
with  $D=4-2\epsilon$ standing for the space-time
dimension. 

In practice $Z_m$ is usually computed from the vector and
scalar parts of the quark self-energy 
$\Sigma_V(p^2)$ and $\Sigma_S(p^2)$ .
In our convention, the bare quark propagator is proportional to 
$\left[\not\!p\left(1+\Sigma_V^0(p^2)\right)
-m_q^0\left(1-\Sigma_S^0(p^2)\right)\right]^{-1}
$. Requiring
the finiteness of the renormalized quark propagator 
and keeping only massless and linear in $m_q$ terms, one  arrives
at  the following recursive   equations  to find $Z_m$ 
\beq
Z_m Z_2 = 1 +  K_\ep
\left\{ 
Z_m Z_2 \Sigma_S^0(p^2)
\right\}
, \ \ \ 
Z_2 =1 - K_\ep
\left\{ 
Z_2  \Sigma_V^0(p^2)
\right\}
\label{Z2Zm}
{},
\eeq
where $K_\epsilon f(\epsilon)$ stands for the singular part of the
Laurent expansion of $f(\epsilon)$ in $\epsilon$ near $\epsilon=0$ and
$Z_2$ is the quark wave function renormalization constant.
Eqs. (\ref{Z2Zm}) express $Z_m$  through massless propagator-type 
(that is dependent on  one external momentum only) 
Feynman integrals (FI), termed as  {\em p-integrals} below.

In this letter  we first present the results of the analytic calculation of
the quark mass anomalous dimension to ${\cal O}(\alpha_s^4)$.  Then we
apply it to evaluate the running of a heavy quark mass in QCD with the
number of active flavours $n_f$ varied from 3 to 6.

\section{Results and discussion}

Taken literally, Eqs. (\ref{Z2Zm}) require calculation of a host
of one-, two-, three-, and {\em four}-loop p-integrals to find $Z_m$ to
${\cal O}(\alpha_s^4)$.  At present a direct analytical calculation of
a p-integral with the loop number not exceeding three is a rather 
easy  business.  First, there exists an elaborated algorithm --- the
method of integration by parts of Ref.~\cite{me81a,me81b} --- which
allows one to analytically evaluate divergent as well as finite parts
of any three-loop p-integral. Second, the algorithm has been neatly
and reliably implemented in the language FORM \cite{Ver91} as the
package named MINCER in Ref.~\cite{mincer2}.

The situation with four-loop diagrams is quite different. At present
there is simply no way to directly compute the divergent part of a
four-loop p-integral.  An indirect and rather involved approach
to perform such calculations analytically is to use the method of
Infrared Rearrangement (IRR) of Ref.~\cite{Vladimirov80} enforced by  
a special technique of dealing with infrared divergences --- the
$R^*$-operation of Refs.~\cite{me84,me91}. Recently, the technique has been
dramatically simplified by explicitly resolving all the relevant
combinatorics for a particular case of 4-loop diagrams contributing to
the  correlator of two  scalar or vector currents in Refs. \cite{gssq} and
\cite{gvvq} respectively. We have extended these improvements for the
problem at hand.

The four-loop diagrams contributing to Eqs. (\ref{Z2Zm}) to order
$\alpha_s^4$ (about 6000) have been generated with the program QGRAF
\cite{qgraf}, then  globally rearranged to a product of
some three-loop p-integrals with a trivial (essentially one-loop)
massive Feynman integral and, finally, computed with the program
MINCER. All calculations have been done in ${\mbox{MS}}$-scheme. As is
well-known, every anomalous dimension is the same for ${\mbox{MS}}$-
and $\overline{\mbox{MS}}$ schemes \cite{MSbar}, thus, the results are
also valid for the latter one.

Our result for the quark mass anomalous dimension reads
(in  evaluating the colour factors we have assumed the case of the
colour group $SU_c(N)$; $n_f$ stands for the number of  quark flavours)
\beq
\g_0 = \frac{N^2-1}{4N} 
\left[
 \frac{3}{2}\right],
\ \ \ \ 
\g_1 = \frac{N^2-1}{16N^2}  
\left\{
-\frac{3}{8} 
+\frac{203}{24} N^2
{+} \,n_f 
\left[
-\frac{5}{6}  N 
\right]
\right\}
\nonumber\\
{},
\eeq
\begin{eqnarray}
\lefteqn{\g_2 = \frac{N^2-1}{64N^3} 
\left\{ \rule{0cm}{6mm} \right.
\frac{129}{16} 
-\frac{129}{16} N^2
+\frac{11413}{216} N^4
}
\nonumber\\
&{+}& \,n_f 
\left[
\frac{23}{4}  N 
-\frac{1177}{108} N^3
-6  N  \,\zeta(3)
-6 N^3 \,\zeta(3)
\right]
{+} \, n_f^2
\left[
-\frac{35}{54} N^2
\right]
\left. \rule{0cm}{6mm} \right\}
{},
\end{eqnarray}
\begin{eqnarray}
\lefteqn{\g_3= \frac{N^2-1}{256N^4} 
\left\{ \rule{0cm}{6mm} \right.
\frac{1261}{128} 
+\frac{50047}{384}  N^2
-\frac{66577}{1152}  N^4
+\frac{460151}{1152}  N^6
+21  \,\zeta(3)}
\nonumber \\ &{}& 
\phantom{+ }
-\frac{47}{2}  N^2 \,\zeta(3)
+52  N^4 \,\zeta(3)
+\frac{1157}{18}  N^6 \,\zeta(3)
-110  N^4 \,\zeta(5)
-110  N^6 \,\zeta(5)
\nonumber\\
&{+}& \,n_f 
\left[
\frac{37}{6}  N 
+\frac{10475}{216}  N^3
-\frac{11908}{81}  N^5
-\frac{111}{2}  N  \,\zeta(3)
-85  N^3 \,\zeta(3)
-\frac{889}{6}  N^5 \,\zeta(3)
 \right. \nonumber \\ &{}& \left.
\phantom{+ \,n_f }
+33  N^3 \,\zeta(4)
+33  N^5 \,\zeta(4)
-30  N  \,\zeta(5)
+50  N^3 \,\zeta(5)
+80  N^5 \,\zeta(5)
\right]
\\
&{+}& \, n_f^2
\left[
-\frac{19}{27}  N^2
+\frac{899}{324}  N^4
+10  N^2 \,\zeta(3)
+10  N^4 \,\zeta(3)
-6  N^2 \,\zeta(4)
-6  N^4 \,\zeta(4)
\right]
\nonumber\\
\nonumber
&{+}& \, n_f^3
\left[
-\frac{83}{162}  N^3
+\frac{8}{9}  N^3 \,\zeta(3)
\right]
\left. \rule{0cm}{6mm} \right\}
{}.
\end{eqnarray}
Here $\zeta$ is the Riemann zeta-function 
($\zeta(3) = 1.202056903\dots$, $\zeta(4) = \pi^4/90$
and  $\zeta(5)  =1.036927755\dots$).
With $N=3$ one  gets the following result for QCD
\beq
\g_0 = 1,
\ \ \ \
\g_1 = \frac{1}{16}
\left\{ \rule{0cm}{6mm} \right.
 \frac{202}{3}
{+} \,n_f 
\left[
-\frac{20}{9}\right]
\left. \rule{0cm}{6mm} \right\}
{},
\eeq
\beq
\g_2 = \frac{1}{64} 
\left\{ \rule{0cm}{6mm} \right.
 1249
{+} \,n_f 
\left[
-\frac{2216}{27} 
-\frac{160}{3}  \,\zeta(3)
\right]
{+} \, n_f^2
\left[
-\frac{140}{81}\right]
\left. \rule{0cm}{6mm} \right\}
{},
\eeq
\begin{eqnarray}
\lefteqn{\g_3= \frac{1}{256} 
\left\{ \rule{0cm}{6mm} \right.
\frac{4603055}{162} 
+\frac{135680}{27}  \,\zeta(3)
-8800  \,\zeta(5)
}
\nonumber\\
&{+}& \,n_f 
\left[
-\frac{91723}{27} 
-\frac{34192}{9}  \,\zeta(3)
+880  \,\zeta(4)
+\frac{18400}{9}  \,\zeta(5)
\right]
\nonumber\\
&{+}& \, n_f^2
\left[
\frac{5242}{243} 
+\frac{800}{9}  \,\zeta(3)
-\frac{160}{3}  \,\zeta(4)
\right]
{+} \, n_f^3
\left[
-\frac{332}{243} 
+\frac{64}{27}  \,\zeta(3)
\right]
\left. \rule{0cm}{6mm} \right\}
{}.
\end{eqnarray}
Note that in three-loop order we exactly reproduce
the known result of Ref.~\cite{Tar82,Larin:massQCD}.
The    four-loop term proportional to 
$n_f^3$ is in agreement to the one found in 
Ref.~\cite{1overNf}.

In  numerical form $\g_m$ reads
\bea
\nonumber
\g_m =  &-& a_s - a_s^2  (4.20833 - 0.138889 n_f)
\\ \nonumber
&-&
a_s^3  (19.5156 - 2.28412 n_f - 0.0270062 n_f^2 )  
\\ \nonumber 
&-&
a_s^4  (98.9434 - 19.1075 n_f + 0.276163 n_f^2  + 0.00579322 n_f^3 )
{}.
\label{N[gm4qcd]}
\eea
It is amusing to compare  the exact result
$\g_3^{\rm exact} = 44.2629\dots$  for $n_f=3$  and   a recent estimation 
of Ref.~ \cite{strange}
\beq
\g^{\rm est}_3(n_f =3) = \frac{\g_2^2(n_f =3)}{\g_1(n_f =3)}
= 40.6843
{}.
\label{gm3est}
\eeq
For the case of QED with $n_f$ identical fermions
our result reads
\beq
\g^{{\rm QED}}_0 = \frac{3}{4},
\ \ \ 
\g^{{\rm QED}}_1 = \frac{1}{16} 
\left\{ \rule{0cm}{6mm} \right.
 \frac{3}{2}
{+} \,n_f 
\left[
-\frac{10}{3}\right]
\left. \rule{0cm}{6mm} \right\}
{},
\eeq
\beq
\g^{{\rm QED}}_2 = \frac{1}{64} 
\left\{ \rule{0cm}{6mm} \right.
 \frac{129}{2}
{+} \,n_f 
\left[
-46 
+48  \,\zeta(3)
\right]
{+} \, n_f^2
\left[
-\frac{140}{27}\right]
\left. \rule{0cm}{6mm} \right\}
{},
\eeq
\begin{eqnarray}
\lefteqn{\g^{{\rm QED}}_3= \frac{1}{256} 
\left\{ \rule{0cm}{6mm} \right.
-\frac{1261}{8} 
-336  \,\zeta(3)
{+} \,n_f 
\left[
-\frac{88}{3} 
+72  \,\zeta(3)
-480  \,\zeta(5)
\right]
}
\nonumber\\
&{+}& \, n_f^2
\left[
\frac{304}{27} 
-160  \,\zeta(3)
+96  \,\zeta(4)
\right]
{+} \, n_f^3
\left[
-\frac{664}{81} 
+\frac{128}{9}  \,\zeta(3)
\right]
\left. \rule{0cm}{6mm} \right\}
{}.
\end{eqnarray}

The RG equation (\ref{anom-mass-def}) is  usually solved by 
\beq
\frac{m(\mu)}{m(\mu_0)} = \frac{c(a_s(\mu))}{c(a_s(\mu_0))}
{},
\eeq
where 
\bea \nonumber
c(x) &=& (x)^{\bar{\g_0}}  \left\{ 1 + (\bar{\g_1} - \bar{\beta_1}\bar{\g_0})x
\right.
\\ \nonumber
&+&
\frac{1}{2}
\left[
(\bar{\g_1} - \bar{\beta_1}\bar{\g_0})^2 
+
\bar{\g_2} + \bar{\beta_1}^2\bar{\g_0}
- \bar{\beta_1}\bar{\g_1} -\bar{\beta_2}\bar{\g_0}
\right] x^2
\\ \label{c(x)}
&+&
\left[ 
\frac{1}{6}(\bar{\g_1} - \bar{\beta_1}\bar{\g_0})^3
+
\frac{1}{2}(\bar{\g_1} - \bar{\beta_1}\bar{\g_0})
(
\bar{\g_2} + \bar{\beta_1}^2\bar{\g_0}
- \bar{\beta_1}\bar{\g_1} -\bar{\beta_2}\bar{\g_0}
)
\right.
\\ \nonumber
&&
+\frac{1}{3}\left(
\left.\left.
\bar{\g_3}
-\bar{\beta_1}^3\bar{\g_0} + 2\bar{\beta_1} \bar{\beta_2}\bar{\g_0}
-\bar{\beta_3}\bar{\g_0} + \bar{\beta_1}\bar{\g_1}
- \bar{\beta_2}\bar{\g_1} - \bar{\beta_1}\bar{\g_2}
\right)
\right] x^3 + {\cal O}(x^4)
\right\}
{}.
\eea
Here $\bar{\g_i} = \g_i/\beta_0$, $\bar{\beta_i} = \beta_i/\beta_0$,
(i=1,2,3) and $\beta_i$ are the coefficient of the QCD beta-function
defined as:
\begin{equation}  \label{a4}
\mu^2\frac{d}{d\mu^2} 
\left( \frac{\alpha_s(\mu )}{\pi} \right)|{{}_{g_{{\rm B}},
 m_{{\rm B}} }}
=
 \beta \equiv
-\sum_{i\geq0}\beta_i\left(\frac{\alpha_s}{\pi}\right)^{i+2}
{}.
 \end{equation}
Equation (\ref{c(x)}) explicitly demonstrates that the knowledge of
the four-loop coefficients $\g_3$ and $\beta_3$ is absolutely necessary
for the self-consistent running of the quark mass in the cases when
the mass-dependent terms of order $\alpha_s^3$ are taken into
account. Examples can be found in Refs.~\cite{gssq,strange,mq2as3}.

The four-loop  beta-function  has been recently  analytically computed
in Ref.~\cite{beta4} with the result
\begin{equation}\label{a8}
\begin{array}{ll}\displaystyle
\beta_0 = &\displaystyle\frac{1}{4}\left(11-\frac{2}{3}n_f\right),
  \;\;\;\;  
\beta_1=\frac{1}{16}\left(102-\frac{38}{3}n_f\right), 
\\[3ex] \displaystyle
\beta_2  = & \displaystyle \frac{1}{64}\left(\frac{2857}{2}
-\frac{5033}{18}n_f+ 
\frac{325}{54}n_f^2\right),
\\[3ex]
\beta_3  = & \dsp\frac{1}{256}\left( 
                 \frac{149753}{6} + 3564\zeta(3)
\right.
\\[3ex]
\rule{0cm}{.4cm} &
\left.
 \dsp
                 -\left[\frac{1078361}{162} + \frac{6508}{27}\zeta(3)\right]n_f
                 +\left[\frac{50065}{162}+ \frac{6472}{81}\zeta(3)\right]n_f^2 
                 + \frac{1093}{729}n_f^3
                  \right)
{}\, .
\end{array} 
\end{equation} 

Now we are in position to evaluate the c-function numerically with 
four-loop accuracy.  For strange, charm, bottom and top it reads:
\bea
c_s(x) &=& x^{4/9} (1  + 0.895062 x + 1.37143 x^2  + 1.95168 x^3,
 \ \ \ (n_f =3),
\nonumber
\\
c_c(x) &=& x^{12/25}(1 + 1.01413 x + 1.38921 x^2  + 1.09054 x^3),
 \ \ \ (n_f =4),
\\
c_b(x) &=& x^{12/23}(1+ 1.17549 x + 1.50071 x^2  + 0.172478 x^3),
 \ \ \ (n_f =5),
\nonumber
\\
\nonumber
c_t(x) &=& x^{4/7}(1+  1.39796 x + 1.79348 x^2  - 0.683433 x^3),
 \ \ \ (n_f =6).
\eea

\vskip0.3cm  

\noindent
{\Large{\bf Acknowledgments}}
\vskip0.3cm

I  would like to thank Matthias Steinhauser for  the support and
careful reading the manuscript.

\end{document}